\documentclass[10pt]{iopart}
\usepackage{iopams}
\usepackage[dvips]{graphicx}
\usepackage{bm}
\usepackage{array}

\renewcommand{\vec}[1]{\bm{#1}}

\newcommand{\ii}{\mathrm{i}}
\newcommand{\dm}{\mathrm{d}}
\newcommand{\pp}{\partial}
\newcommand{\nablabf}{\bm{\nabla}}
\newcommand{\rhoe}{\rho_\mathrm{el}}
\newcommand{\lamD}{\lambda_\mathrm{D}}
\newcommand{\kBT}{k_\mathrm{B}T}
\newcommand{\Matlab}{\textsc{Matlab}}
\newcommand{\Comsol}{\textsc{Comsol}}

\begin{document}

\title[Gregersen et al.: Topology and shape optimization of ICEO micropumps]{Topology and shape optimization of induced-charge electro-osmotic micropumps}

\author{M M Gregersen$^1$, F Okkels$^1$, M Z Bazant$^{2,3}$, and
H Bruus$^1$}

\address{$^1$Department of Micro- and Nanotechnology, Technical University of Denmark,\\
DTU Nanotech, Building 345 East, DK-2800 Kongens Lyngby, Denmark\\
$^2$Departments of Chemical Engineering and Mathematics,\\
MIT, Cambridge, MA 02139 USA\\
$^3$Physico-Chimie Th\'eorique, UMR 7083 Gulliver-CNRS,\\ ESPCI, 10
rue Vauquelin, Paris 75005, France}
\ead{Henrik.Bruus@nanotech.dtu.dk \hfill 13 January 2009}

\begin{abstract}
For a dielectric solid surrounded by an electrolyte and positioned
inside an externally biased parallel-plate capacitor, we study
numerically how the resulting induced-charge electro-osmotic (ICEO)
flow depends on the topology and shape of the dielectric solid. In
particular, we extend existing conventional electrokinetic models
with an artificial design field to describe the transition from the
liquid electrolyte to the solid dielectric. Using this design field,
we have succeeded in applying the method of topology optimization to
find system geometries with non-trivial topologies that maximize the
net induced electro-osmotic flow rate through the electrolytic
capacitor in the direction parallel to the capacitor plates. Once
found, the performance of the topology optimized geometries has been
validated by transferring them to conventional electrokinetic models
not relying on the artificial design field. Our results show the
importance of the topology and shape of the dielectric solid in ICEO
systems and point to new designs of ICEO micropumps with
significantly improved performance.
\end{abstract}


\submitto{\NJP}


\section{Introduction}
\label{sec:intro}

Induced-charge electro-osmotic (ICEO) flow is generated when an
external electric field polarizes a solid object placed in an
electrolytic solution~\cite{Bazant2004a,Levitan2005}. Initially, the
object acquires a position-dependent potential difference $\zeta$
relative to the bulk electrolyte. However, this potential is
screened out by the counter ions in the electrolyte by the formation
of an electric double layer of width $\lamD$ at the surface of the
object.  The ions in the diffusive part of the double layer are then
electromigrating in the resulting external electric field, and by
viscous forces they drag the fluid along. At the outer surface of
the double layer a resulting effective slip velocity is thus
established. For a review of ICEO see Squires and
Bazant~\cite{Squires2004}.

The ICEO effect may be utilized in microfluidic devices for fluid
manipulation, as proposed in 2004 by Bazant and
squires~\cite{Bazant2004a}. Theoretically, various simple dielectric
shapes have been analyzed analytically for their ability to pump and
mix liquids~\cite{Squires2004,Squires2006}. Experimentally ICEO was
observed and the basic model validated against particle image
velocimetry in 2005~\cite{Levitan2005}, and later it has been used
in a microfluidic mixer, where a number of triangular shapes act as
passive mixers \cite{Harnett2008}. However, no studies have been
carried out concerning the impact of topology changes of the
dielectric shapes on the mixing or pumping efficiency. In this work
we focus on the application of topology optimization to ICEO
systems. With this method it is possible to optimize the dielectric
shapes for many purposes, such as mixing and pumping efficiency.

Our model system consists of two externally biased, parallel
capacitor plates confining an electrolyte. A dielectric solid is
shaped and positioned in the electrolyte, and the external bias
induces ICEO flow at the dielectric surfaces. In this work we focus
on optimizing the topology and shape of the dielectric solid to
generate the maximal flow perpendicular to the external applied
electric field. This example of establishing an optimized ICEO
micropump serves as demonstration of the implemented topology
optimization method.

Following the method of Borrvall and Petersson~\cite{Borrvall2003}
and the implementation by Olesen, Okkels and Bruus~\cite{Olesen2006}
of topology optimization in microfluidic systems we introduce an
artificial design field $\gamma(\vec{r})$ in the governing
equations. The design field varies continuously from zero to unity,
and it defines to which degree a point in the design domain is
occupied by dielectric solid or electrolytic fluid. Here, $\gamma =
0$ is the limit of pure solid and $\gamma = 1$ is the limit of pure
fluid, while intermediate values of $\gamma$ represent a mixture of
solid and fluid. In this way, the discrete problem of placing and
shaping the dielectric solid in the electrolytic fluid is converted
into a continuous problem, where the sharp borders between solid and
electrolyte are replaced by continuous transitions throughout the
design domain. In some sense one can think of the solid/fluid
mixture as a sort of ion-exchange membrane in the form of a sponge
with varying permeability.  This continuum formulation allows for an
efficient gradient-based optimization of the problem.

In one important aspect our system differs from other systems
previously studied by topology optimization: induced-charge
electro-osmosis is a boundary effect relying on the polarization and
screening charges in a nanometer-sized region around the solid/fluid
interface. Previously studied systems have all been relying on bulk
properties such as the distribution of solids in mechanical stress
analysis~\cite{Bendsoe2003}, photonic band gap structures in optical
wave guides~\cite{Jensen2004}, and acoustic noise
reduction~\cite{Duhring2008}, or on the distribution of solids and
liquids in viscous channel
networks~\cite{Borrvall2003,Olesen2006,GersborgHansen2005} and
chemical microreactors~\cite{Okkels2007}. In our case, as for most
other applications of topology optimization, no mathematical proof
exists that the topology optimization routine indeed will result in
an optimized structure. Moreover, since the boundary effects of our
problem result in a numerical analysis which is very sensitive on
the initial conditions, on the meshing, and on the specific form of
the design field, we take the more pragmatic approach of finding
non-trivial geometries by use of topology optimization, and then
validate the optimality by transferring the geometries to
conventional electrokinetic models not relying on the artificial
design field.

\section{Model system}
\begin{figure}[t]
 \centering
 \includegraphics[width=0.85\linewidth,clip]{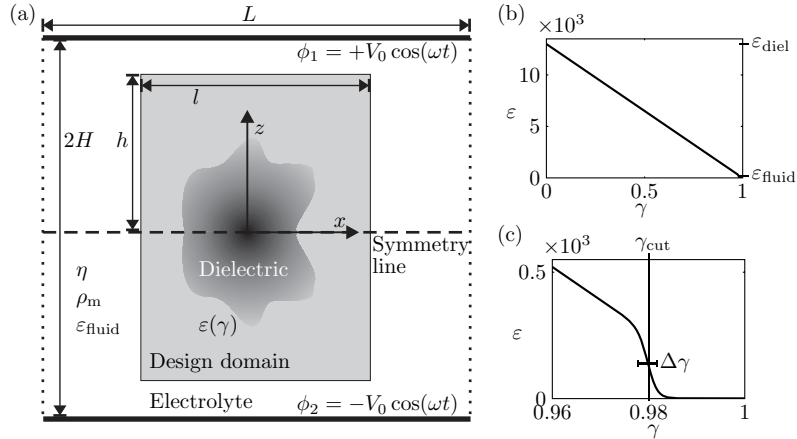}
\caption{\label{fig:ModelSystem} (a) A sketch of the rectangular
$L\times2H$ cross-section of the electrolytic capacitor in the
$xz$-plane. The external voltage $\phi_1$ and $\phi_2$ is applied to
the two infinite parallel-plate electrodes (thick black lines) at
$z=\pm H$. The voltage difference $\phi_1-\phi_2$ induces an ICEO
flow around the un-biased dielectric solid (dark gray) shaped by the
topology optimization routine limited to the rectangular $l\times
2h$ design domain (light gray). The dielectric solid is surrounded
by pure electrolyte (light gray and white). Periodic boundary
conditions are applied ar the vertical edges (dotted lines). (b) The
dimensionless electric permittivity $\varepsilon$ as a function of
the design variable $\gamma$. (c) Zoom-in on the rapid convergence
of $\varepsilon(\gamma)$ towards $\varepsilon_\mathrm{fluid}=1$ for
$\gamma$ approaching unity after passing the value
$\gamma_\mathrm{cut}-\Delta\gamma \simeq 0.98$.}
\end{figure}

We consider a parallel-plate capacitor externally biased with a
harmonic oscillating voltage difference $\Delta\phi =2V_0\cos(\omega
t)$ and enclosing an electrolyte and a dielectric solid. The two
capacitor plates are positioned at $z=\pm H$ and periodic boundary
conditions are applied at the open ends at $x=\pm L/2$. The
resulting  bound domain of size $L\times 2H$  in the $xz$-plane is
shown in Fig.~\ref{fig:ModelSystem}. The system is assumed to be
unbounded and translational invariant in the perpendicular
$y$-direction. The topology and shape of the dielectric solid is
determined by the numerical optimization routine acting only within
a smaller rectangular, central design domain of size $l\times 2h$.
The remaining domain outside this area is filled with pure
electrolyte.  Double layers, or Debye screening layers, are formed
in the electrolyte around each piece of dielectric solid to screen
out the induced polarization charges. The pull from the external
electric field on these screening layers in the design domain drives
an ICEO flow in the entire domain.

If the dielectric solid is symmetric around the $x$-axis, the
anti-symmetry of the applied external bias voltage ensures that the
resulting electric potential is anti-symmetric and the velocity and
pressure fields symmetric around the center plane $z=0$. This
symmetry appears in most of the cases studied in this paper, and
when present it is exploited to obtain a significant decrease in
memory requirements of the numerical calculations.

The specific goal of our analysis is to determine the topology and
shape of the dielectric solid such that a maximal flow rate $Q$ is
obtained parallel to the $x$-axis, i.e.\ perpendicular to the
direction of external potential field gradient.

\section{Governing equations}
We follow the conventional continuum approach to the electrokinetic
modeling of the electrolytic capacitor~\cite{Squires2004}. For
simplicity we consider a symmetric, binary electrolyte, where the
positive and negative ions with concentrations $c_+$ and $c_-$,
respectively, have the same diffusivity $D$ and valence charge
number $Z$.

\subsection{Bulk equations in the conventional ICEO model}
\label{sec:BulkEquations}
Neglecting chemical reactions in the bulk of the electrolyte, the
ionic transport is governed by particle conservation through the
continuity equation,
\begin{equation}\label{eq:particlecons}
\frac{\pp c_\pm}{\pp t} = -\nablabf\cdot\vec{J}_\pm,
\end{equation}
where $\vec{J}_\pm$ is the flux density of the two ionic species,
respectively. Assuming a dilute electrolytic solution, the ion flux
densities are governed by the Nernst--Planck equation,
\begin{equation}\label{eq:NP_gov}
\vec{J}_\pm = -D\bigg(\nablabf c_\pm +\frac{\pm Ze}{\kBT}c_\pm\nablabf\phi\bigg),
\end{equation}
where the first term expresses ionic diffusion and the second term
ionic electro-migration due to the electrostatic potential $\phi$.
Here $e$ is the elementary charge, $T$ the absolute temperature and
$k_\mathrm{B}$ the Boltzmann constant. We note that due to the low
fluid velocity $\vec{v}$ obtained in the ICEO systems under
consideration, we can safely neglect the convective ion fluxes
$c_\pm\vec{v}$ throughout this paper, see
Table~\ref{tab:CalcQuantities}.

The electrostatic potential $\phi$ is determined by the charge density
$\rhoe = Ze(c_+ - c_-)$ through Poisson's equation,
 \begin{equation}
 \label{eq:phi1_gov} \nablabf
 \cdot(\varepsilon_\mathrm{fluid}\nablabf\phi) = -\rhoe,
 \end{equation}
where $\varepsilon_\mathrm{fluid}$ is the fluid permittivity, which
is assumed constant. The fluid velocity field $\vec{v}$ and pressure
field $p$ are governed the the continuity equation and the
Navier--Stokes equation for incompressible fluids,
 \numparts
 \begin{eqnarray}
 \label{eq:ContEq_gov}
 \nablabf\cdot\vec{v} = 0,\\
 \rho_\mathrm{m}
 \bigg[\frac{\pp\vec{v}}{\pp t}+(\vec{v}\cdot\nablabf)\vec{v}\bigg]
 = -\nablabf p + \eta \nabla^2\vec{v} - \rhoe\nablabf\phi,
 \label{eq:NS_gov}
 \end{eqnarray}
 \endnumparts
where $\rho_\mathrm{m}$ and $\eta$ are the fluid mass density and
viscosity, respectively, both assumed constant.

\subsection{The artificial design field $\gamma$ used in the topology optimization model of ICEO}
To be able to apply the method of topology optimization, it is
necessary to extend the conventional ICEO model with three
additional terms, all functions of a position-dependent artificial
design field $\gamma(\vec{r})$.  The design field varies
continuously from zero to unity, where $\gamma = 0$ is the limit of
a pure dielectric solid and $\gamma = 1$ is the limit of a pure
electrolytic fluid. The intermediate values of $\gamma$ represent a
mixture of solid and fluid.

The first additional term concerns the purely fluid dynamic part of
our problem. Here, we follow Borrvall and
Petersson~\cite{Borrvall2003} and model the dielectric solid as a
porous medium giving rise to a Darcy friction force density
$-\alpha(\gamma)\vec{v}$, where $\alpha(\gamma)$ may be regarded as
a local inverse permeability, which we denote the Darcy friction. We
let $\alpha(\gamma)$ be a linear function of  $\gamma$ of the form
$\alpha(\gamma)=\alpha_\mathrm{max}(1-\gamma)$, where
$\alpha_\mathrm{max} = \eta/\ell_\mathrm{pore}^2$ is the Darcy
friction of the porous dielectric material assuming a characteristic
pore size $\ell_\mathrm{pore}$. In the limit of a completely
impenetrable solid the value of $\alpha_\mathrm{max}$ approaches
infinity, which leads to a vanishing fluid velocity $\vec{v}$. The
modified Navier--Stokes equation extending to the entire domain,
including the dielectric material, becomes
 \begin{equation}
 \label{eq:NS_gov_gamma}
 \rho_\mathrm{m}
 \bigg[\frac{\pp\vec{v}}{\pp t}+(\vec{v}\cdot\nablabf)\vec{v}\bigg]
 = -\nablabf p +\eta \nabla^2\vec{v} -\rhoe\nablabf\phi -\alpha(\gamma)\vec{v}.
 \end{equation}

The second additional term is specific to our problem. Since the
Navier--Stokes equation is now extended to include also the porous
dielectric medium, our model must prevent the unphysical penetration
of the electrolytic ions into the solid. Following the approach of
Kilic \textit{et al.}~\cite{Killic2007}, where current densities are
expressed as gradients of chemical potentials, $\vec{J} \propto
-\nablabf\mu$, we model the ion expulsion by adding an extra free
energy term $\kappa(\gamma)$ to the chemical potential $\mu_\pm =\pm
Ze\phi +\kBT \ln(c_\pm/c_0)+\kappa(\gamma)$ of the ions, where $c_0$
is the bulk ionic concentration for both ionic species. As above we
let $\kappa(\gamma)$ be a linear function of $\gamma$ of the form
$\kappa(\gamma) = \kappa_\mathrm{max}(1-\gamma)$, where
$\kappa_\mathrm{max}$ is the extra energy cost for an ion to enter a
point containing a pure dielectric solid as compared to a pure
electrolytic fluid. The value of $\kappa_\mathrm{max}$ is set to an
appropriately high value to expel the ions efficiently from the
porous material while still ensuring a smooth transition from
dielectric solid to electrolytic fluid. The modified ion flux
density becomes
\begin{equation}\label{eq:NP_gov_gamma}
\vec{J}_\pm = -D\bigg(\nablabf c_\pm +\frac{\pm Ze}{\kBT}c_\pm\nablabf\phi
+\frac{1}{\kBT}c_\pm\nablabf\kappa(\gamma)\bigg).
\end{equation}

The third and final additional term is also specific to our problem.
Electrostatically, the transition from the dielectric solid to the
electrolytic fluid is described through the Poisson equation by a
$\gamma$-dependent permittivity $\varepsilon(\gamma)$. This modified
permittivity varies continuously between the value
$\varepsilon_\mathrm{diel}$ of the dielectric solid and
$\varepsilon_\mathrm{fluid}$ of the electrolytic fluid. As above, we
would like to choose $\varepsilon(\gamma)$ to be a linear function
of $\gamma$. However, during our analysis using a metallic
dielectric with $\varepsilon_\mathrm{diel} = 10^6\varepsilon_0$ in
an aqueous electrolyte with $\varepsilon_\mathrm{fluid} =
78\:\varepsilon_0$ we found unphysical polarization phenomena in the
electrolyte due to numerical rounding-off errors for $\gamma$ near,
but not equal to, unity. To overcome this problem we ensured a more
rapid convergence towards the value $\varepsilon_\mathrm{fluid}$ by
introducing a cut-off value $\gamma_\mathrm{cut} \simeq 0.98$, a
transition width $\Delta\gamma \simeq 0.002$, and the following
expression for $\varepsilon(\gamma)$,
\begin{equation}
\varepsilon(\gamma) =
\varepsilon_\mathrm{diel}+(\varepsilon_\mathrm{fluid}-\varepsilon_\mathrm{diel})
\bigg\{1-\frac{(1\!-\!\gamma)}{2}
\Big[\tanh\!\Big(\frac{\gamma_\mathrm{cut}\!-\!\gamma}{\Delta\gamma}\Big)
+1\Big]\bigg\}.
\end{equation}
For $\gamma \lesssim \gamma_\mathrm{cut}$ we obtain the linear
relation $\varepsilon(\gamma) = \varepsilon_\mathrm{diel} +
(\varepsilon_\mathrm{fluid}-\varepsilon_\mathrm{diel})\gamma$, while
for $\gamma \gtrsim \gamma_\mathrm{cut}$ we have
$\varepsilon(\gamma) = \varepsilon_\mathrm{fluid}$, see
Fig.~\ref{fig:ModelSystem}(b)-(c). For $\gamma$ sufficiently close
to unity (and not only when $\gamma$ equals unity with numerical
precision), this cut-off procedure ensures that the calculated
topological break up of the dielectric solid indeed leads to several
correctly polarized solids separated by pure electrolyte. The
modified Poisson equation becomes
 \begin{equation}\label{eq:phi1_gov_gamma}
 \nablabf\cdot\big[\varepsilon(\gamma)\nablabf\phi\big] = -\rhoe.
 \end{equation}

Finally, we introduce the $\gamma$-dependent quantity, the so-called
objective function $\Phi[\gamma]$, to be optimized by the topology
optimization routine: the flow rate in the $x$-direction
perpendicular to the applied potential gradient. Due to
incompressibility, the flow rate $Q(x)$ is the same through
cross-sections parallel to the $yz$-plane at any position $x$. Hence
we can use the numerically more stable integrated flow rate as the
objective function,
 \begin{equation} \label{eq:ObjFunc}
 \Phi[\gamma(\vec{r})] =\int_0^L Q(x)\: \dm x =\int_\Omega\vec{v}\cdot\hat{\vec{n}}_x\: \dm x\:\dm z,
 \end{equation}
where $\Omega$ is the entire geometric domain (including the design
domain), and $\hat{\vec{n}}_x$ the unit vector in the $x$ direction.

\subsection{Dimensionless form}
To prepare the numerical implementation, the governing equations are
rewritten in dimensionless form, using the characteristic parameters
of the system. In conventional ICEO systems the size $a$ of the
dielectric solid is the natural choice for the characteristic length
scale $\ell_0$, since the generated slip velocity at the solid
surface is proportional to $a$. However, when employing topology
optimization we have no prior knowledge of this length scale, and
thus we choose it to be the fixed geometric half-length $\ell_0 = H$
between the capacitor plates. Further characteristic parameters are
the ionic concentration $c_0$ of the bulk electrolyte, and the
thermal voltage $\phi_0=\kBT/(Ze)$. The characteristic velocity
$u_0$ is chosen as the Helmholtz--Smoluchowski slip velocity induced
by the local electric field $E=\phi_0/\ell_0$, and finally the
pressure scale is set by the characteristic microfluidic pressure
scale $p_0=\eta u_0/\ell_0$.

Although strictly applicable only to parallel-plate capacitors, the
characteristic time $\tau_0$ of the general system in chosen as the
RC time of the double layer in terms of the Debye length $\lamD$ of
the electrolyte~\cite{Bazant2004b},
 \begin{equation}
 \tau_0= \frac{\ell_0}{D}\:\lamD
 = \frac{\ell_0}{D}\: \sqrt{\frac{\kBT\varepsilon_\mathrm{fluid}}{2(Ze)^2c_0}}.
 \end{equation}

Moreover, three characteristic numbers are connected to the
$\gamma$-dependent terms in the governing equations: The
characteristic free energy $\kappa_0$, the characteristic
permittivity chosen as the bulk permittivity
$\varepsilon_\mathrm{fluid}$, and the characteristic Darcy friction
coefficient $\alpha_0$. In summary,
\numparts
\begin{equation}
\ell_0 = H, \qquad \phi_0=\frac{\kBT}{Ze}, \quad
u_0=\frac{\varepsilon_\mathrm{fluid}\phi_0^2}{\eta\,\ell_0}, \quad
p_0=\frac{\eta\,u_0}{\ell_0},
\end{equation}
\begin{equation}
\tau_0=\frac{\ell_0\lamD}{D}, \quad\; \omega = \frac{2\pi}{\tau_0},
\quad\;\;\; \kappa_0=\kBT, \qquad \alpha_0=\frac{\eta}{\ell_0^2}.
\end{equation}
\endnumparts
The new dimensionless variables (denoted by a tilde) thus become
\numparts
\begin{equation}
\tilde{\vec{r}}=\frac{\vec{r}}{\ell_0}, \quad
\tilde{\vec{v}}=\frac{\vec{v}}{u_0}, \quad
\tilde{p}=\frac{p}{p_0}, \quad \;
\tilde{\phi}=\frac{\phi}{\phi_0}, \quad
\tilde{c}_\pm=\frac{c_\pm}{c_0},
\end{equation}
\begin{equation}
\tilde{t}=\frac{t}{\tau_0}, \quad \;
\tilde{\kappa}=\frac{\kappa}{\kappa_0}, \quad
\tilde{\alpha}= \frac{\alpha}{\alpha_0}, \quad
\tilde{\varepsilon}= \frac{\varepsilon}{\varepsilon_\mathrm{fluid}}.
\end{equation}
\endnumparts
In the following all variables and parameters are made dimensionless
using these characteristic numbers and for convenience the tilde is
henceforth omitted.

\subsection{Linearized and reformulated equations}
To reduce the otherwise very time and memory consuming numerical
simulations, we choose to linearize the equations. There are several
nonlinearities to consider.

By virtue of a low Reynolds number $\textit{Re} \approx 10^{-6}$,
see Table~\ref{tab:CalcQuantities}, the nonlinear Navier--Stokes
equation is replaced by the linear Stokes equation. Likewise, as
mentioned in Sec.~\ref{sec:BulkEquations}, the low P\'eclet number
$\textit{P\'e}\approx 10^{-3}$ allows us to neglect the nonlinear
ionic convection flux density $c_\pm\vec{v}$. This approximation
implies the additional simplification that the electrodynamic
problem is independent of the hydrodynamics.

Finally, since $Ze\zeta/\kBT \lesssim 0.5$ the linear
Debye--H{\"u}ckel approximation is valid, and we can utilize that
the ionic concentrations only deviate slightly from the bulk
equilibrium ionic concentration. The governing equations are
reformulated in terms of the average ion concentration $c\equiv
(c_++c_-)/2$ and half the charge density $\rho\equiv (c_+-c_-)/2$.
Thus, by expanding the fields to first order as $c=1+\delta c$ and
$\rho=0+\delta\rho$, the resulting differential equation for $\rho$
is decoupled from that of $c$. Introducing complex field notation,
the applied external bias voltage is $\Delta\phi(t)=2V_0\cos(\omega
t) =\mathrm{Re}[2V_0\exp(\ii\omega t)]$, yielding a corresponding
response for the potential $\phi$ and charge density $\rho$, with
the complex amplitudes
$\Phi(\vec{r})=\Phi_R(\vec{r})+\ii\Phi_I(\vec{r})$ and
$P(\vec{r})=P_R(\vec{r})+\ii P_I(\vec{r})$, respectively. The
resulting governing equations for the electrodynamic problem is then
\numparts
\begin{eqnarray}
\qquad \qquad \quad
\nablabf\cdot\big[\varepsilon(\gamma)\nablabf\Phi_R\big] &= -\frac{1}{\epsilon^2}P_R,\\
\qquad \qquad \quad
\nablabf\cdot\big[\varepsilon(\gamma)\nablabf\Phi_I\big] &= -\frac{1}{\epsilon^2}P_I,\\
\nablabf\cdot\big[\nablabf\Phi_R +\nablabf P_R +P_R\nablabf\kappa(\gamma)\big] &= -\frac{\omega}{\epsilon}P_I,\\
\nablabf\cdot\big[\nablabf\Phi_I +\nablabf P_I +P_I\nablabf\kappa(\gamma)\big] &= +\frac{\omega}{\epsilon}P_R,
\end{eqnarray}
\endnumparts
where we have introduced the dimensionless thickness of the linear
Debye layer $\epsilon=\lamD/\ell_0$. Given the electric potential
$\Phi$ and the charge density $P$, we solve for the time-averaged
hydrodynamic fields $\langle\vec{v}\rangle$ and $\langle p\rangle$,
\numparts
\begin{eqnarray}
\nablabf\cdot\langle\vec{v}\rangle &= 0,\\
\qquad 0 &= -\nablabf\langle p\rangle + \nabla^2
\langle\vec{v}\rangle + \langle \vec{f}_\mathrm{el}\rangle -
\alpha(\gamma)\langle\vec{v}\rangle,
\end{eqnarray}
where the time-averaged electric body force density $\langle
\vec{f}_\mathrm{el}\rangle$ is given by
 \begin{equation}
 \quad
 \langle \vec{f}_\mathrm{el}\rangle =
 -\frac{1}{2\epsilon^2}\big[P_R\nablabf\Phi_R +P_I\nablabf\Phi_I\big].
 \end{equation}
\endnumparts

\subsection{Boundary conditions}
For symmetric dielectric solids we exploit the symmetry around $z=0$
and consider only the upper half $(0<z<1)$ of the domain. As
boundary condition on the driving electrode we set the amplitude
$V_0$ of the applied potential. Neglecting any electrode reactions
taking place at the surface there is no net ion flux in the normal
direction to the boundary with unit vector $\hat{\vec{n}}$. For the
fluid velocity we set a no-slip condition, and thus at $z=1$ we have
\numparts
\begin{equation}
\Phi_R = V_0 , \quad \Phi_I = 0,
\end{equation}
\begin{equation}
\hat{\vec{n}}\cdot\big[\nablabf\Phi_R +\nablabf P_R +P_R\nablabf\kappa(\gamma)\big] = 0,
\end{equation}
\begin{equation}
\hat{\vec{n}}\cdot\big[\nablabf\Phi_I +\nablabf P_I +P_I\nablabf\kappa(\gamma)\big] = 0,
\end{equation}
\begin{equation}
\langle\vec{v}\rangle = \bm{0}.
\end{equation}
\endnumparts
On the symmetry axis ($z=0$) the potential and the charge density
must be zero due to the anti-symmetry of the applied potential.
Furthermore, there is no fluid flux in the
normal direction and the shear stresses vanish. So at $x=0$ we have
\numparts
\begin{equation}
\Phi_R = \Phi_I = 0, \quad
P_R = P_I = 0,
\end{equation}
\begin{equation}
\hat{\vec{n}}\cdot\langle\vec{v}\rangle = 0, \quad
\hat{\vec{t}}\cdot\langle\vec{\sigma}\rangle\cdot\hat{\vec{n}}=0,
\end{equation}
\endnumparts
where the dimensionless stress tensor is $\langle\sigma_{ik}\rangle=
-\langle p\rangle\delta_{ik} +\big(\partial_i\langle v_k\rangle
+\partial_k\langle v_i\rangle\big)$, and $\hat{\vec{n}}$ and
$\hat{\vec{t}}$ are the normal and tangential unit vectors,
respectively, where the latter in 2D, contrary to 3D, is uniquely
defined. On the remaining vertical boundaries ($x=\pm L/2\ell_0$)
periodic boundary conditions are applied to all the fields.

Corresponding boundary conditions apply to the conventional ICEO
model Eqs.~(\ref{eq:particlecons})-(\ref{eq:NS_gov}), without the
artificial design field but with a hard-wall dielectric solid. For
the boundary between a dielectric solid and and electrolytic fluid
the standard electrostatic conditions apply, moreover, there is no
ion flux normal to the surface, and a no-slip condition is applied
to the fluid velocity.

\begin{table}[!b]
\caption{\label{tab:parameters} Parameters used in the simulations
of the topology optimization ICEO model and the conventional ICEO
model.} \centering
\begin{tabular}{l*{3}{>{$}c<{$}}}
\hline
Parameter                 &  \textrm{Symbol}  & \textrm{Dimensionless} &  \textrm{Physical} \\
& & \textrm{value} & \textrm{value} \\
\hline
Characteristic length     &   \ell_0 & 1.0     & 250~\mathrm{nm} \\
Channel half-height       &    H     & 1.0     & 250~\mathrm{nm} \\
Channel length            &    L     & 2.0     & 500~\mathrm{nm} \\
Design domain half-height &    h     & 0.8     & 200~\mathrm{nm} \\
Design domain length      &    l     & 0.6     & 150~\mathrm{nm} \\
Linear Debye length       &  \lamD   & 0.08    & 20~\mathrm{nm} \\
Characteristic velocity   &  u_0     & 1.0     & 1.7\times 10^{-3}~\mathrm{m/s} \\
Characteristic potential  & \phi_0   & 1.0     & 25~\mathrm{mV} \\
External potential amplitude & V_0   & 1.0     & 25~\mathrm{mV} \\
External potential frequency & \omega & 6.28 & 4\times 10^5~\mathrm{rad/s} \\
Bulk fluid permittivity   & \varepsilon_\mathrm{fluid} & 1.0 & 78\:\varepsilon_0 \\
Dielectric permittivity   & \varepsilon_\mathrm{diel}  & 1.3\times 10^4 & 10^6\:\varepsilon_0 \\
Bulk ionic concentration  &  c_0     & 1.0      & 0.23~\mathrm{mM} \\
Fluid viscosity           &  \eta    & 1.0      & 10^{-3}~\mathrm{Pa\,s} \\
Ionic diffusion constant  &   D      & 1.0      & 2\times 10^{-9}\mathrm{m}^2/\mathrm{s} \\
Ionic free energy in solid & \kappa        & 3.0  & 75~\mathrm{mV} \\
Maximum Darcy friction  & \alpha_\mathrm{max}  & 10^{5}  &  2\times 10^{16}~\mathrm{Pa\:s/m}^{2} \\
\hline
\end{tabular}
\end{table}

\section{Implementation and validation of numerical code}
\subsection{Implementation and parameter values}
For our numerical analysis we use the commercial numerical
finite-element modeling tool \Comsol~\cite{COMSOL} controlled by
scripts written in \Matlab~\cite{MATLAB}. The mathematical method of
topology optimization in microfluidic systems is based on the
seminal paper by Borrvall and Petersson~\cite{Borrvall2003}, while
the implementation containing the method of moving asymptotes by
Svanberg~\cite{Svanberg1987,SvanbergMMA} is taken from Olesen,
Okkels and Bruus~\cite{Olesen2006}.

Due to the challenges discussed in Sec.~\ref{sec:ValidConventional}
of resolving all length scales in the electrokinetic model, we have
chosen to study small systems, $2H = 500$~nm, with a relatively
large Debye length, $\lamD = 20$~nm. Our main goal is to provide a
proof of concept for the use of topology optimization in
electro-hydrodynamic systems, so the difficulties in fabricating
actual devices on this sub-micrometric scale is not a major concern
for us in the present work. A list of the parameter values chosen
for our simulations is given in Table~\ref{tab:parameters}.

For a typical topology optimization, as the one shown in
Fig.~\ref{fig:topopt_result}(a), approximately 5400 FEM elements are
involved. In each iteration loop of the topology optimization
routine three problems are solved: the electric problem , the
hydrodynamic problem, and the adjunct problem for the sensitivity
analysis, involving $4\times10^4$ , $2\times10^4$ , and
$7\times10^4$ degrees of freedom, respectively. On an Intel Core 2
Duo 2~GHz processer with 2~GB RAM the typical CPU time is several
hours.

\subsection{Analytical and numerical validation by the conventional ICEO model}
\label{sec:ValidConventional}

We have validated our simulations in two steps. First, the
conventional ICEO model not involving the design field
$\gamma(\vec{r})$ is validated against the analytical result for the
slip velocity at a simple dielectric cylinder in an infinite AC
electric field given by Squires and Bazant \cite{Squires2004}.
Second, the design field model is compared to the conventional
model. This two-step validation procedure is necessary because of
the limited computer capacity. The involved length scales in the
problem make a large number of mesh elements necessary for the
numerical solution by the finite element method. Four different
length scales appear in the gamma-dependent model for the problem of
a cylinder placed mid between the parallel capacitor plates: The
distance $H$ from the center of the dielectric cylinder to the
biased plates, the radius $a$ of the cylinder, the Debye length
$\lamD$, and the length $d$ over which the design field $\gamma$
changes from zero to unity. This last and smallest length-scale $d$
in the problem is controlled directly be the numerical mesh size set
up in the finite element method. It has to be significantly smaller
than $\lamD$ to model the double layer dynamics correctly, so here a
maximum value of the numerical mesh size is defined.

The analytical solution of Squires and Bazant~\cite{Squires2004} is
only strictly valid in the case of an infinitely thin Debye layer in
an infinite electric field. So, to compare this model with the
bounded numerical model the plate distance must be increased to
minimize the influence on the effective slip velocity. Furthermore,
it has been shown in a numerical study of Gregersen \textit{et
al.}~\cite{Gregersen2009} that the Debye length $\lamD$ should be
about a factor of $10^3$ smaller than the cylinder radius $a$ to
approximate the solution for the infinitely thin Debye layer model.
Including the demand of $d$ being significantly smaller than $\lamD$
we end up with a length scale difference of at least $10^5$, which
is difficult to resolve by the finite element mesh, even when mesh
adaption is used. Consequently, we have checked that the slip
velocity for the conventional model converges towards the analytical
value when the ratio $\lamD/a$ decreases. Afterwards, we have
compared the solutions for the conventional and gamma-dependent
models in a smaller system with a ratio of $\lamD/a\sim 10$ and
found good agreement.

\subsection{Validation of the self-consistency of the topology optimization} \label{sec:ValidSelfConsist}

\begin{figure}[t]
 \centering
 \includegraphics[height=50mm,clip]{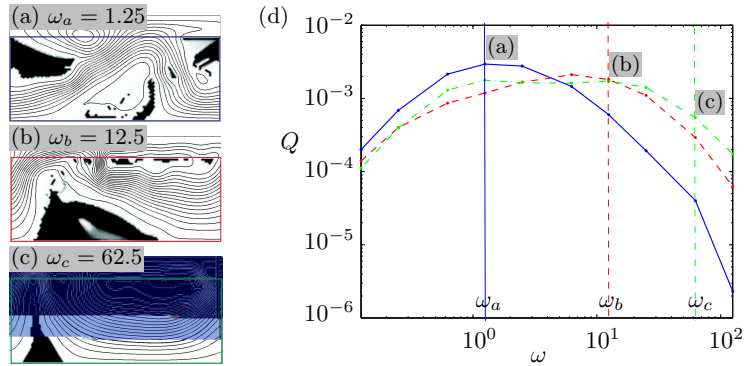}
\caption{\label{fig:QvsOmega} Validation of the self-consistency of
the topology optimization for different driving frequencies
$\omega$. (a) The streamline pattern (thick lines) for $\omega =
\omega_a = 1.25$ calculated using the design-field ICEO model with a
porous dielectric medium (black and gray), the structure $\gamma_a$
of which has been found by topology optimization within the
indicated rectangular design domain (straight lines). The flow rate
for this converged solution structure is $Q=2.95\times 10^{-3}$. (b)
As panel (a) but with $\omega = \omega_b = 12.5$ and $Q=1.82\times
10^{-3}$. (c) As panel (a) but with $\omega = \omega_c = 62.5$ and
$Q=0.55\times 10^{-3}$. (d) Flow rate $Q$ versus frequency $\omega$
for each of the three structures in panel (a), (b), and (c). Note
that structure $\gamma_a$ indeed yields the highest flow rate $Q$
for $\omega = \omega_a$, structure $\gamma_b$ maximizes $Q$ for
$\omega = \omega_b$, and structure $\gamma_c$ maximizes $Q$ for
$\omega = \omega_c$.}
\end{figure}

As an example of our validation of the self-consistency of the
topology optimization, we study the dependence of the objective
function $Q = \Phi[\omega,\gamma(\omega,\vec{r})]$ on the external
driving frequency $\omega$. As shown in
Fig.~\ref{fig:QvsOmega}(a)-(c) we have calculated the topology
optimized dielectric structures $\gamma_j =
\gamma(\omega_j,\vec{r})$, $j=a,b,c$, for three increasing
frequencies $\omega = \omega_a = 1.25$, $\omega = \omega_b = 12.5$,
and $\omega = \omega_c = 62.5$. In the following we let
$Q_j(\omega)$ denote the flow rate calculated at the frequency
$\omega$ for a structure optimized at the frequency $\omega_j$.

First, we note that $Q_j(\omega_j)$ decreases as the frequency
increases above the characteristic frequency $\omega_0 = 1$;
$Q_a(\omega_a) = 2.95\times10^{-3}$, $Q_b(\omega_b) =
1.82\times10^{-3}$, and $Q_c(\omega_c) = 0.55\times10^{-3}$. This
phenomenon is a general aspect of ICEO systems, where the largest
effect is expected to happen at $\omega = 2\pi/\tau_0 = 6.28$.

Second, and most significant, we see in Fig.~\ref{fig:QvsOmega}(d)
that structure $\gamma_a$ is indeed the optimal structure for
$\omega = \omega_a$ since $Q_a(\omega_a) > Q_b(\omega_a),
Q_c(\omega_a)$. Likewise, $\gamma_b$ is optimal for $\omega =
\omega_b$, and $\gamma_c$ is optimal for $\omega = \omega_c$.

We have gained confidence in the self-consistency of our topology
optimization routine by carrying out a number of tests like the one
in the example above.

\section{Results}
\begin{figure}[t]
 \centering
 \includegraphics[clip,width=\linewidth]{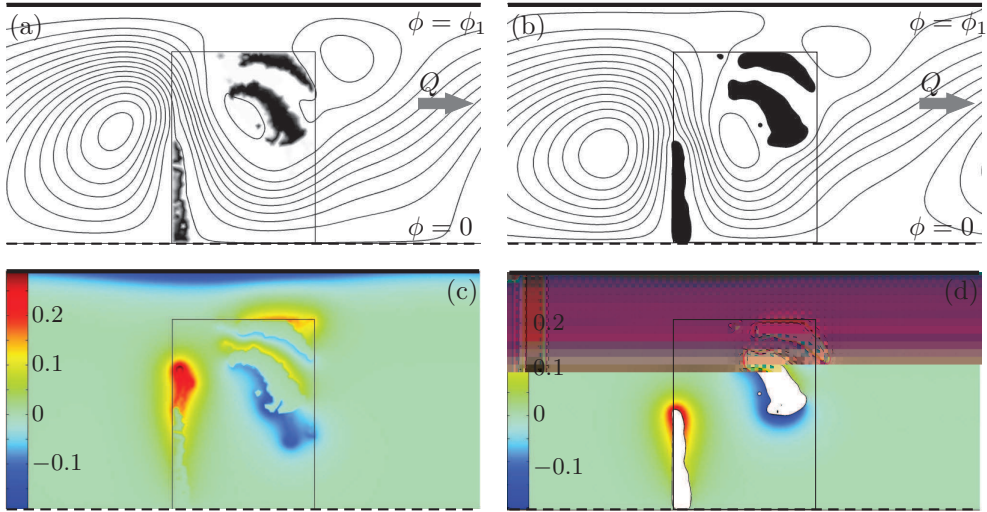}
\caption{\label{fig:topopt_result} (a) The streamline pattern (thick
lines) calculated for $\omega=6.28$ using the design-field ICEO
model with a porous dielectric medium (black and gray), the
structure of which has been find by topology optimization within the
indicated rectangular design domain (thin lines). The flow rate for
this converged solution structure is $Q=2.99\times 10^{-3}$. (b) The
streamline pattern (full lines) calculated using the conventional
ICEO model with a hard-walled dielectric solid (black). The shape of
the dielectric solid is the 0.95-contour of the $\gamma$-field taken
from the topology-optimized structure shown in panel (a). The flow
rate is $Q=Q^*= 1.88\times 10^{-3}$. (c) and (d) Color plot of the
charge density $\rho(\vec{r})$ corresponding to panel (a) and (b),
respectively. See Table~\ref{tab:parameters} for parameter values.}
\end{figure}

\subsection{Topology optimization}

For each choice of parameters the topology optimization routine
converges to a specific distribution of dielectric solid given by
$\gamma(\vec{r})$. As a starting point for the investigation of the
optimization results we used the parameters listed in
Table~\ref{tab:parameters}. As discussed above, the geometric
dimensions are chosen as large as possible within the computational
limitations: the Debye length is set to $\lamD = 20~\mathrm{nm}$ and
the distance between the capacitor plates to $2H = 500~\mathrm{nm}$.
The external bias voltage is of the order of the thermal voltage
$V_0 = 25$~mV to ensure the validity of the linear Debye--H\"{u}ckel
approximation. We let the bulk fluid consist of water containing
small ions, such as dissolved KCl, with a concentration $c_0 =
0.23$~mM set by the chosen Debye length. The dielectric material
permittivity is set to $\varepsilon_\mathrm{diel}=
10^{6}\:\varepsilon_0$ in order to mimic the characteristics of a
metal. The artificial parameters $\kappa$ and $\alpha_\mathrm{max}$
are chosen on a pure computational basis, where they have to mimic
the real physics in the limits of fluid and solid, but also support
the optimization routine when the phases are mixed.

\begin{table}[b]
\caption{\label{tab:CalcQuantities} The value of characteristic
physical quantities calculated in the topology optimization ICEO
model corresponding to Fig.~\ref{fig:topopt_result}.} \centering
\begin{tabular}{l*{3}{>{$}c<{$}}}
\hline
Quantity                 &  \textrm{Symbol}  & \textrm{Dimensionless} &  \textrm{Physical} \\
& & \textrm{value} & \textrm{value} \\
\hline
Gap between dielectric pieces &   \ell_\mathrm{gap} & 0.4     & 100~\mathrm{nm} \\
Velocity in the gap       &    u_\mathrm{gap}     &   0.016\:u_0   & 28~\mu\mathrm{m/s} \\
Largest zeta potential    &    \zeta_\mathrm{max} & 0.5\:\phi_0     & 12.5~\mathrm{mV} \\
Reynolds number \textit{Re}  &     \rho_\mathrm{m} u_\mathrm{gap} \ell_\mathrm{gap}/\eta     & 2.8\times 10^{-6}     & -\\
P\'eclet number \textit{P\'e} &    u_\mathrm{gap}\ell_\mathrm{gap}/D    & 1.4\times 10^{-3}     & - \\
Debye--H\"{u}ckel number \textit{H\"{u}} &  e\zeta_\mathrm{max}/(4\kBT) & 0.13   & - \\
\hline
\end{tabular}
\end{table}

Throughout our simulations we have primarily varied the applied
frequency $\omega$ and the size $l\times 2h$ of the design domain.
In Fig.~\ref{fig:QvsOmega} we have shown examples of large design
domains with $l\times h = 2.0 \times 0.8$ covering 80\% of the
entire domain and frequency sweeps over three orders of magnitude.
However, in the following we fix the frequency to be $\omega =
2\pi/\tau_0 = 6.28$, where the ICEO response is maximum. Moreover we
focus on a smaller design domain $l\times h = 0.6 \times 0.8$ to
obtain better spatial resolution for the given amount of computer
memory and to avoid getting stuck in local optima. It should be
stressed that the size of the design domain has a large effect on
the specific form and size of the dielectric islands produced by the
topology optimization. Also, it is important if the design domain is
allowed to connect to the capacitor plates or not, see the remarks
in Sec.~\ref{sec:conclusion}.

The converged solution found by topology optimization under these
conditions is shown in Fig.~\ref{fig:topopt_result}(a). The shape of
the porous dielectric material is shown together with a streamline
plot of equidistant contours of the flow rate. We notice that many
stream lines extend all the way through the domain from left to
right indicating that a horizontal flow parallel to the $x$-axis is
indeed established. The resulting flow rate is
$Q=2.99\times10^{-3}$. The ICEO flow of this solution, based on the
design-field model, is validated by transferring the geometrical
shape of the porous dielectric medium into a conventional ICEO model
with a hard-walled dielectric not depending on the design field. In
the latter model the sharp interface between the dielectric solid
and the electrolyte is defined by the 0.95-contour of the topology
optimized design field $\gamma(\vec{r})$. The resulting geometry and
streamline pattern of the conventional ICEO model is shown in
Fig.~\ref{fig:topopt_result}(b). The flow rate is now found to be
$Q=Q^*= 1.88\times10^{-3}$. There is a close resemblance between the
results of two models both qualitatively and quantitatively. It is
noticed how the number and positions of the induced flow rolls match
well, and also the absolute values of the induced horizontal flow
rates differs only by 37\%.

Based on the simulation we can now justify the linearization of our
model. The largest velocity $u_\mathrm{gap}$ is found in the gap of
width $\ell_\mathrm{gap}$ between the two satellite pieces and the
central piece. As listed in Table~\ref{tab:CalcQuantities} the
resulting Reynolds number is $\textit{Re} = 2.8\times 10^{-7}$, the
P\'eclet number is $\textit{P\'e} = 1.4\times 10^{-3}$, while the
Debye--H\"{u}ckel number is $\textit{H\"{u}} = 0.13$.

\begin{figure}[b]
 \centering
 \includegraphics[clip,width=\linewidth]{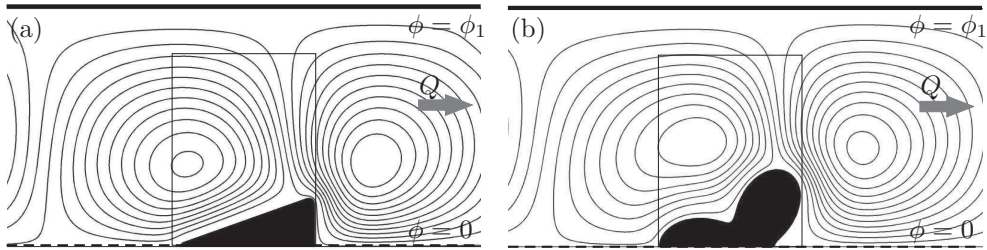}
\caption{\label{fig:ref_struc} (a) The streamline pattern (thick
lines) for a simple triangular reference structure calculated for
$\omega=6.28$ using the conventional ICEO model with a hard-walled
dielectric solid (black). The height $b=0.32$  of the triangle is
chosen to give the largest flow rate for a fixed base line given by
the rectangular design domain of Fig.~\ref{fig:topopt_result}(a).
The flow rate is $Q=0.22\times 10^{-3}$. (b) The same as panel (a)
except the geometry of the dielectric solid is given by the
perturbed circle $r(\theta) = 0.24[1+0.5\cos(3\theta)]$. The flow
rate is $Q=0.46\times 10^{-3}$.}
\end{figure}

\subsection{Comparison to simple shapes}
We evaluate our result for the optimized flow rate by comparing it
to those obtained for more basic, simply connected, dielectric
shapes, such as triangles and perturbed circles previously studied
in the literature as flow inducing objects both analytically and
experimentally \cite{Squires2004,Squires2006,Harnett2008}. For
comparison, such simpler shapes have been fitted into the same
design domain as used in the topology optimization
Fig.~\ref{fig:topopt_result}(a), and the conventional ICEO model
without the design field was solved for the same parameter set. In
Fig.~\ref{fig:ref_struc}(a) the resulting flow for a triangle with
straight faces and rounded corners is shown. The height $b$ of the
face perpendicular to the symmetry line was varied within the height
of the design domain $0<b<0.8$, and the height $b=0.32$ generating
the largest flow in the $x$-direction results in a flow rate of
$Q=0.22\times10^{-3}$, which is eight times smaller than the
topology optimized result. In Fig.~\ref{fig:ref_struc}(b) the
induced flow around a perturbed cylinder with radius
$r(\theta)=0.24\big[1+0.5\cos(3\theta)\big]$ is depicted. Again the
shape has been fitted within the allowed design domain. The
resulting flow rate $Q=0.46\times10^{-3}$ is higher than for the
triangle but still a factor of four slower than the optimized
result. It is clearly advantageous to change the topology of the
dielectric solid from simply to multiply connected.

For the topology optimized shape in Fig.~\ref{fig:topopt_result}(a)
it is noticed that only a small amount of flow is passing between
the two closely placed dielectric islands in the upper left corner
of the design domain. To investigate the importance of this
separation, the gap between the islands was filled out with
dielectric material and the flow calculated. It turns out that this
topological change only lowered the flow rate slightly ($15\%$) to a
value of $Q=1.59\times10^{-3}$. Thus, the important topology of the
dielectric solid in the top-half domain is the appearance of one
center island crossing the antisymmetry line and one satellite
island near the tip of the center island.

\subsection{Shape optimization}
The topology optimized solutions are found based on the extended
ICEO model involving the artificial design field $\gamma(\vec{r})$.
To avoid the artificial design field it is desirable to validate and
further investigate the obtained topology optimized results by the
physically more correct conventional ICEO model involving
hard-walled solid dielectrics. We therefore extend the reasoning
behind the comparison of the two models shown in
Fig.~\ref{fig:topopt_result} and apply a more stringent shape
optimization to the various topologies presented above. With this
approach we are gaining further understanding of the specific shapes
comprising the overall topology of the dielectric solid. Moremore,
it is possible to point out simpler shapes, which are easier to
fabricate, but still perform well.

In shape optimization the goal is to optimize the objective function
$\Phi$, which depends on the position and shape of the boundary
between the dielectric solid and the electrolytic fluid. This
boundary is given by a line interpolation through a small number of
points on the boundary. These control points are in turn given by
$N$ design variables $\vec{g} = (g_1, g_2, \ldots, g_N)$, so the
objective function of Eq.~(\ref{eq:ObjFunc}) depending on the design
field $\gamma(\vec{r})$ is now written as $\Phi[\vec{g}]$ depending
on the design variables $\vec{g}$,
 \begin{equation} \label{eq:ObjFuncShape}
 \Phi[\vec{g}] = \int_\Omega\vec{v}\cdot\hat{\vec{n}}_x\: \dm x\:\dm z.
 \end{equation}

To carry out the shape optimization we use a direct bounded
Nelder-Mead simplex method \cite{Nelder1965} implemented in
\Matlab~\cite{Lagarias1998,Matlab_fminsearch}. This robust method
finds the optimal point $\vec{g}_\mathrm{opt}$ in the
$N$-dimensional design variable space by initially creating a
simplex in this space, e.g.\ a $N$-dimensional polyhedron spanned by
$N+1$ points, one of which is the initial guess. The simplex then
iteratively moves towards the optimal point by updating one of the
$N+1$ points at the time. During the iteration, the simplex can
expand along favorable directions, shrink towards the best point, or
have its worst point replaced with the point obtained by reflecting
the worst point through the centroid of the remaining $N$ points.
The iteration terminates once the extension of the simplex is below
a given tolerance. We note that unlike topology optimization, the
simplex method relies only on values of the objective function
$\Phi[\vec{g}]$ and not on the sensitivity $\partial \Phi/\partial
\vec{g}$~\cite{unbound}.

\begin{figure}[t]
\centering
\includegraphics[clip,width=\linewidth]{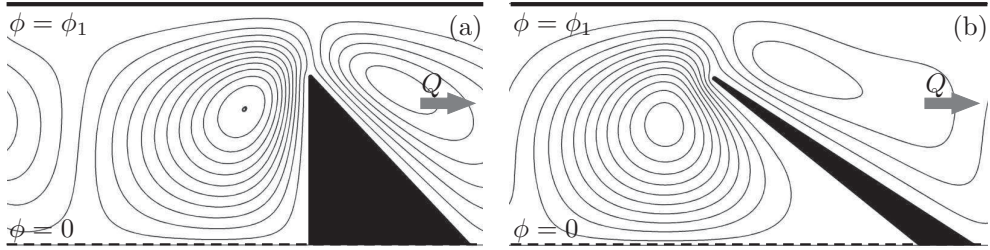}
\caption{\label{fig:ShapeOpt_triangle} (a) The streamline pattern
(thick lines) for the shape-optimized right-angled triangle fixed at
the symmetry line $z=0$ calculated for $\omega=6.28$ using the
conventional ICEO model with a hard-walled dielectric solid (black).
In the full domain this is a triangle symmetric around $z=0$. The
flow rate is $Q=0.32\times 10^{-3}$. (b) As in panel (a) but without
constraining the triangle to be right-angled. In the full domain the
shape is foursided polygon symmetric around $z=0$. The flow rate is
$Q=0.76\times 10^{-3}$. Note that all sharp corners of the polygons
have been rounded by circular arcs of radius 0.01.}
\end{figure}

First, we perform shape optimization on a right-angled triangle
corresponding to the one shown in  Fig.~\ref{fig:ref_struc}(a). Due
to the translation invariance in the $x$-direction, we fix the first
basepoint of the triangle $(x_1,0)$ to the right end of the
simulation domain, while the second point $(x_2,0)$ can move freely
along the baseline, in contrast to the original rectangular design.
To ensure a right-angled triangle only the $z$-coordinate of the top
point $(x_2,z_2)$ may move freely. In this case the design variable
becomes the two-component vector $\vec{g} = (x_2,z_2)$. The optimal
right-angled triangle is shown in
Fig.~\ref{fig:ShapeOpt_triangle}(a). The flow rate is $Q=0.32\times
10^{-3}$ or 1.5 times larger than that of the original right-angled
triangle confined to the design domain.

If we do not constrain the triangle to be right-angled, we instead
optimize a polygon shape spanned by three corner points in the upper
half of the electrolytic capacitor. So, due to the symmetry of the
problem, we are in fact searching for the most optimal, symmetric
foursided polygon. The three corner points are now given as
$(x_1,0)$, $(x_2,0)$,and $(x_3,z_3)$, and again due to translation
invariance, it results in a three-component design variable $\vec{g}
= (x_2,x_3,z_3)$. The resulting shape-optimized  polygon is shown in
Fig.~\ref{fig:ShapeOpt_triangle}(b). The flow rate is $Q=0.76\times
10^{-3}$, which is 3.5 times larger than that of the original
right-angled triangle confined to the design domain and 2.4 times
better than that of the best right-angled triangle. However, this
flow rate is still a factor of 0.4 lower than the topology optimized
results.

\begin{figure} [t]
 \centering
 \includegraphics[height=50mm,clip]{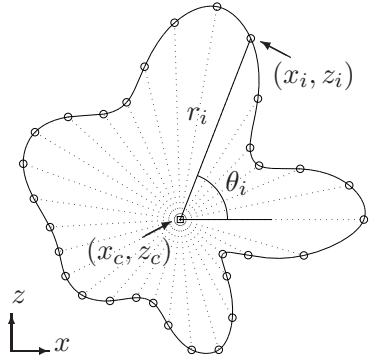}
 \caption{\label{fig:ShapeOpt_Setup} Illustration of the
parametrization, Eq.~(\ref{eq:ShapeDescript}), of the boundary of a
dielectric solid with a complex shape. The polar representation is
shown for point $i=7$. The shape consist of five harmonic components
represented by Eq.~(\ref{eq:HarmModes}) with the design-variables
$x_c=-0.1312$, $z_c=0.7176$, $r_0=0.1403$, $A_i=\{0.2501, 0.0151,
0.0062, 0.2103, 0.2313\}$, $\varphi_i = \{ -1.7508, -2.2526, 0.4173,
0.1172, -0.2419\}$.}
 \end{figure}

To be able to shape optimize the more complex shapes of
Fig.~\ref{fig:topopt_result} we have employed two methods to obtain
a suitable set of design variables. The first method, the radial
method, is illustrated in Fig.~\ref{fig:ShapeOpt_Setup}. The
boundary of a given dielectric solid is defined through a cubic
interpolation line through $N$ control points $(x_i,z_i),
i=1,2,\ldots,N$, which are parameterized in terms of two
co-ordinates $(x_c,z_c)$ of a center point, two global scale factors
$A$ and $B$, $N$ lengths $r_i$, and $N$ fixed angles $\theta_i$
distributed in the interval from 0 to $2\pi$,
\begin{equation}
 \label{eq:ShapeDescript}
 (x_i,z_i) = (x_c,z_c) + r_i\: (A\cos\theta_i,\: B\sin\theta_i).
\end{equation}
In this case the design variable becomes $\vec{g} = (x_c,z_c,r_1,r_2,\ldots,r_N,A,B)$.

The second parametrization method involves a decomposition into
harmonic components. As before we define a central point $(x_c,z_c)$
surrounded by $N$ control points. However now, the distances $r_i$
are decomposed into $M$ harmonic components given by
 \begin{equation}
 \label{eq:HarmModes}
 r_i = r_0 \bigg(1+\sum_{n=1}^M A_n\cos(n\theta_i + \varphi_n)\bigg),
 \end{equation}
where $r_0$ is an overall scale parameter and $\varphi_n$ is a phase
shift. In this case the design variable becomes $\vec{g} = (x_c,z_c,
r_0, A_1, A_2, \ldots,A_M, \varphi_1, \varphi_2,\ldots,\varphi_M)$.

\subsection{Comparing topology optimization and shape optimization}

When shape-optimizing a geometry similar to the one found by
topology optimization, we let the geometry consist of two pieces:
(\textit{i}) an elliptic island centered on the symmetry-axis and
fixed to the right side of design domain, and (\textit{ii}) an
island with a complex shape to be placed anywhere inside the design
domain, but not overlapping with the elliptic island. For the
ellipse we only need to specify the major axis $A$ and the minor
axis $B$, so these two design parameters add to the design variable
listed above for either the radial model or the harmonic
decomposition model. To be able to compare with the topology
optimized solution the dielectric solid is restricted to the design
domain.

\begin{figure} [t]
\centering
\includegraphics[clip,width=\linewidth]{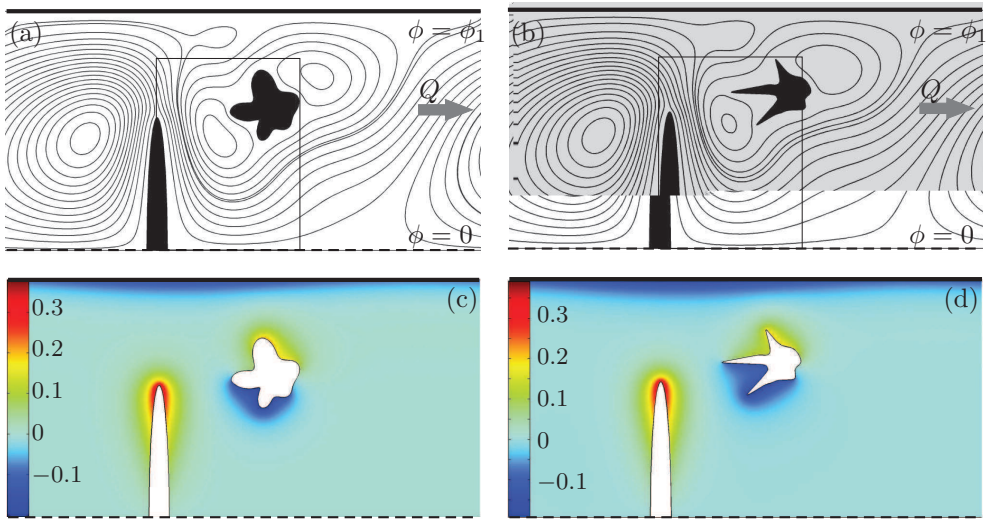}
\caption{\label{fig:ShapeOpt} Shape-optimized dielectrics with a
topology corresponding to the topology-optimized shapes of
Fig.~\ref{fig:topopt_result}. (a) The streamline pattern (thick
lines) for a two-piece geometry calculated using the conventional
ICEO model. The shape of the hard-walled dielectric solid (black) is
found by shape optimization using the radial method
Eq.~(\ref{eq:ShapeDescript}) with $N=16$ directional angles. The
flow rate is $Q=1.92\times 10^{-3}$. (b) The same as panel (a)
except the geometry of the dielectric solid is by shape optimization
using the harmonic decomposition method Eq.~(\ref{eq:HarmModes})
with $M=5$ modes. The flow rate is $Q=1.52\times 10^{-3}$. (c) and
(d) Color plot of the charge density $\rho(\vec{r})$ corresponding
to panel (a) and (b), respectively.}
\end{figure}

The result of this two-piece shape optimization is shown in
Fig.~\ref{fig:ShapeOpt}. Compared to the simply connected
topologies, the two-piece shape-optimized systems yields much
improved flow rates. For the shape optimization involving the radial
method with 16 directional angles and $A=B$ for the complex piece,
the flow rate is $Q=1.92\times 10^{-3}$, Fig.~\ref{fig:ShapeOpt}(a),
which is 2.5 times larger than that of the shape-optimized foursided
symmetric polygon. The harmonic decomposition method,
Fig.~\ref{fig:ShapeOpt}(b), yields a flow rate of $Q=1.52\times
10^{-3}$ or 2.0 times larger than that of the polygon.

All the results for the obtained flow rates are summarized in
Table~\ref{tab:flow_rates}. It is seen that two-piece shape
optimized systems performs as good as the topology optimized system,
when analyzed using the conventional ICEO model without the
artificial design field. We also note by comparing
Figs.~\ref{fig:topopt_result} and~\ref{fig:ShapeOpt} that the
resulting geometry found using either topology optimization or shape
optimization is essentially the same. The central island of the
dielectric solid is a thin structure perpendicular to the symmetry
axis and covering approximately 60\% of the channel width. The
satellite island of complex shape is situated near the tip of the
central island. It has two peaks pointing towards the central island
that seem to suspend a flow roll which guides the ICEO flow through
the gap between the two islands.

\begin{table}[b]
\caption{Overview of the resulting flow rates $Q$ relative to the
topology optimized value $Q^* = 1.88\times10^{3}$, see
Fig.~\ref{fig:topopt_result}(b), for the various geometries analyzed
in the conventional ICEO model. The methods by which the geometries
have been determined are listed.}\label{tab:flow_rates} \centering
\begin{tabular}{l l c}
\hline
Shape                        &   Method      & Flow rate\\
&& $Q/Q^*$ \\
\hline
Triangle with optimal height, Fig.~\ref{fig:ref_struc}(a)
                             &    Shape optimization   & 0.12 \\
Perturbed cylinder, Fig.~\ref{fig:ref_struc}(b)
                             &    Fixed shape          & 0.24 \\
Optimized triangle, Fig.~\ref{fig:ShapeOpt_triangle}(a)
                             & Shape optimization    & 0.17 \\
Optimized foursided polygon, Fig.~\ref{fig:ShapeOpt_triangle}(b)
                             & Shape optimization    & 0.40 \\
\hline
Topology optimized result, Fig.~\ref{fig:topopt_result}(b)
                             & Topology optimization & 1.00 \\
\hline
Harmonic decomposition and ellipse, Fig.~\ref{fig:ShapeOpt}(a)
                             & Shape optimization    & 0.81 \\
Radial varying points and ellipse, Fig.~\ref{fig:ShapeOpt}(b)
                             & Shape optimization    & 1.02 \\
\hline
\end{tabular}
\end{table}

\section{Concluding remarks}
\label{sec:conclusion}
The main result of this work is the establishment of the topology
optimization method for ICEO models extended with the design field
$\gamma(\vec{r})$. In contrast to the conventional ICEO model with
its sharply defined, impenetrable dielectric solids, the design
field ensures a continuous transition between the porous dielectric
solid and the electrolytic fluid, which allows for an efficient
gradient-based optimization of the problem. By concrete examples we
have shown how the use of topology optimization has led to
non-trivial system geometries with a flow rate increase of nearly
one order of magnitude.

However, there exist many local optima, and we cannot be sure that
the converged solution is the global optimum. The resulting shapes
and the generated flow rates depend on the initial condition for the
artificial $\gamma$-field. Generally, the initial condition used
throughout this paper, $\gamma=0.99$ in the entire design domain,
leads to the most optimal results compared to other initial
conditions. This initial value corresponds to a very weak change
from the electrolytic capacitor completely void of dielectric solid.
In contrast, if we let $\gamma=0.01$ corresponding to almost pure
dielectric material in the entire design region, the resulting
shapes are less optimal, i.e.\ the topology optimization routine is
more likely to be caught caught in a local optimum. Furthermore, the
resulting shapes turns out to be mesh-dependent as well. So, we
cannot conclude much about global optima. Instead, we can use the
topology optimized shapes as inspiration to improve existing
designs. For this purpose shape optimization turns out to be a
powerful tool. We have shown in this work how shape optimization can
be used efficiently to refine the shape of the individual pieces of
the dielectric solid once its topology has been identified by
topology optimization.

For all three additional $\gamma$-dependent fields $\alpha(\gamma)$,
$\kappa(\gamma)$, and $\varepsilon(\gamma)$ we have used (nearly)
linear functions. In many previous applications of topology
optimization non-linear functions have successfully been used to
find global optima by gradually changing the non-linearity into
strict linearity during the iterative procedure
\cite{Borrvall2003,Olesen2006,Bendsoe2003,Okkels2007}. However, we
did not improve our search for a global optimum by employing such
schemes, and simply applied the (nearly) linear functions during the
entire iteration process.

The limited size of the design domain is in some cases restricting
the free formation of the optimized structures. This may be avoided
by enlarging the design domain. However, starting a topology
optimization in a very large domain gives a huge amount of degrees
of freedoms, and the routine is easily caught in local minima. These
local minima often yield results not as optimal as those obtained
for the smaller design boxes. A solution could be to increase the
design domain during the optimization iteration procedure. It should
be noted that increasing the box all the way up to the capacitor
plates results in solution shapes, where some of the dielectric
material is attached to the electrode in order to extend the
electrode into the capacitor and thereby maximize the electric field
locally. This may be a desirable feature for some purposes. In this
work we have deliberately avoided such solutions by keeping the
edges of the design domain from the capacitor plates.

Throughout the paper we have only presented results obtained for
dielectric solids shapes forced to be symmetric around the center
plane $z=0$. However, we have performed both topology optimization
and shape optimization of systems not restricted to this symmetry.
In general we find that the symmetric shapes always are good
candidates for the optimal design. It cannot be excluded, though,
that in some cases a spontaneous symmetry breaking occurs similar to
the asymmetric S-turn channel studied in Ref.~\cite{Olesen2006}.

By studying the optimized shapes of the dielectric solids, we have
noted that pointed features often occurs, such as those clearly seen
on the dielectric satellite island in Fig.~\ref{fig:ShapeOpt}(b).
The reason for these to appear seems to be that the pointed regions
of the dielectric surfaces can support large gradients in the
electric potential and associated with this also with large charge
densities. As a result large electric body forces act on the
electrolyte in these regions. At the same time the surface between
the pointed features curve inward which lowers the viscous flow
resistance due to the no-slip boundary condition. This effect is
similar to that obtained by creating electrode arrays of different
heights in AC electro-osmosis~\cite{Bazant2006,Urbanski2006}.

Another noteworthy aspect of the topology optimized structures is
that the appearance of dielectric satellite islands seem to break up
flow rolls that would otherwise be present and not contribute to the
flow rate. This leads to a larger net flow rate, as can be seen be
comparing Figs.~\ref{fig:ref_struc} and~\ref{fig:ShapeOpt}.

Throughout the paper have treated the design field $\gamma$ as an
artificial field. However, the design-field model could perhaps
achieve physical applications to systems containing ion exchange
membranes, as briefly mentioned in Sec.~\ref{sec:intro}. Such
membranes are indeed porous structures permeated by an electrolyte.

In conclusion, our analysis points out the great potential for
improving actual ICEO-based devices by changing simply connected
topologies and simple shapes of the dielectric solids, into multiply
connected topologies of complex shapes.

\section*{Acknowledgement}
We would like to thank Elie Rapha\"{e}l and Patrick Tabeling at
ESPCI for their hospitality and support during our collaboration.


\begin{thebibliography}{99}
\bibitem{unbound} The \Matlab{} function \texttt{fminsearch} is
actually unbounded. Therefore the arcus tangent function is used to
map the unbounded real axis to the bounded interval of the design
variables.
\end{thebibliography}
\end{document}